# Galactic Distribution of Chirality Sources of Organic Molecules

Daniel S. Helman

Faculty of Labor Relations and Trade Unions, Ton Duc Thang University, Ho Chi Minh City, Vietnam
helmandaniel@tdt.edu.vn

Abstract
Conceptualizing planetary habitability depends on understanding how living organisms originated and what features of environments are essential to foster abiogenesis. Estimates of the abundance of life's building blocks are confounded by incomplete knowledge of the role of chirality and racemization in organic compounds in the origination of living organisms. Chirality is an essential feature of enzymes as well as many lock-and-key type structures. There are four known processes that can act on complex organic molecules to promote racemization for abiogenesis: quantum-tunneling effects; selection via interaction with circularly polarized light (CPL); templating processes; and interactions with electrical and magnetic (EM) fields. These occur in different places, respectively: cold interstellar space; regions of space with energetic photons, dust and/or magnetic fields; and mineral surfaces (for both templating and EM fields). Chirality as a feature of terrestrial life suggests neither a special place for local development of homochirality nor for extra-terrestrial enrichment and delivery. The presence of these molecules in three competing scenarios for life's origin—chemical gardens, geothermal fields, and ice substrates—relies on a framework of hypothesis and estimation. An easily-modified worksheet is included in the supplemental material that allows a user to generate different scenarios and data related to the distribution of chiral organic molecules as building blocks for living organisms within the galaxy. A simple hypothetical mechanism for planetary magnetic field reversals, based on a high-density plasma inner core, is also presented as a means to aid in estimating field polarity and hence the orientation of racemization processes based on planetary magnetic fields.

# 1 Introduction

The term *chirality* was introduced by Lord Kelvin in 1893 to refer to geometric figures or groups of points whose mirror image cannot be rotated to attain the original image [1]. One can think of handedness: a right hand is chiral, as no amount of rotation of its mirror image (a left hand) will attain the orientation of the original.

Take a certain Cartesian representation of a hand with coordinates {x, y, z} that one may find or construct for computer animation. An *inversion* about the origin would give the new coordinates {−x, −y, −z} and even if this new form were rotated about an axis, this new form could not be translated (i.e. moved across) to overlap the original completely. These are left-hand and right-hand varieties. They are not equivalent, even under *rotoinversion*. This property is called chirality.

Chiral forms are abundant in mineral species and depend on the symmetry of the unit cell of the crystal. Quartz, for example, has a trigonal unit cell (with a three-per-rotation symmetry axis and three two-per-rotation symmetry axes perpendicular to it, denoted *32* or $D_3$) linked in a network of $SiO_4$ helices that spiral either clockwise or anticlockwise depending on the sample. Eleven of the thirty two *symmetry classes* of minerals are chiral, and many common minerals on the Earth's surface and crust are chiral: quartz, serpentine, epsomite, nepheline and many others.

Many individual molecules as well can exist in chiral forms, and an atom (or group of atoms) that is chemically bonded with four other atoms or groups of atoms will necessarily have chiral forms—since an anisometric four-atom linkage about a center can be arranged in at least two different ways in three-dimensional space. Metals have high valence and can act as such chiral centers. Likewise, the carbon group of elements have valence four, and carbon is notable in forming chiral centers in many compounds, with the chiral centers commonly denoted as *asymmetric carbon atoms* [2].

*Chiral excess* refers to an abundance of one chiral form over another in a sample. The process of chiral enrichment is called *racemization*. The distribution of quartz in nature does not exhibit a chiral excess of either chiral form in the terrestrial environment [3], while the distributions of chiral organic molecules exhibit a striking chiral excess, in particular in living organisms. L-amino acids and D-carbohydrates are the rule in biology, with few exceptions. (In the bacterial cell wall one can find D-amino acids which the bacterium develops to survive from degrading enzymes targeting L-amino acids. This is an adaptive exception.) Yet the abundance of amino acids in living systems is nearly completely *homochiral*.

We usually follow one of two hypotheses to explain this bias—a *conventional hypothesis* which posits a shift from prebiotic equal amounts of both forms, to a non-linear biological process where one form is preferred due to some unknown event. Alternatively, a *deterministic hypothesis* attempts to explain the preference only by natural laws. It is not now clear which hypothesis of homochirality in terrestrial life is likely valid, yet if living organisms are discovered elsewhere in the galaxy following the opposite abundance from what is seen on Earth in an environment that is essentially similar, the deterministic hypothesis would be falsified. An equivalent question is whether the chirality of amino acids and carbohydrates observed in terrestrial life forms is an essential feature of living organisms.

In the Earth's planetary system, two distinct sources of organic matter are likewise hypothesized. Each depends on a different type of environment within distinct types of dust cloud. First, the formation of complex organic compounds on ice and dust grains within the proto-stellar *hot cores* of dust during star formation are evidenced by observation of other systems as well as from experimental results (e.g. showing ethanol and dimethyl ether formation from simple methanol ices) [4], and by computational modeling (e.g. showing formation of the amino acid glycine plus related molecules such as glycinal, propionic acid, and propanal) [5]. Second, hot cores and stellar planetary systems also receive a flux of material from the dark cold (10 K) interior of molecular clouds as well as from the *diffuse interstellar medium*, which itself shows evidence of the presence of complex organic compounds (i.e. in a strong stable spectroscopic signature at 2200Å, and in diffuse interstellar emissions bands of longer wavelength in the signatures of stars, and also of emission and absorption spectra—all currently of unknown origin that are consistent with the presence of complex organic molecules such as polycyclic aromatic hydrocarbons (PAHs), fullerenes, diamonds, or carbonaceous solids, e.g. kerogen compounds. [6] The interactions of these two bodies of material would account for the physical presence of organic compounds in early terrestrial environments as well as in cometary and meteoritic reservoirs of complex organic compounds.

Under the domain of astrobiology, a better understanding of various possible sources of chirality of organic molecules is important to estimate habitability [7]. The word *habitability* has as its root the Latin *habitare*, to possess or live in, and in its planetary-science meaning is a function of several parameters including the presence of liquid water. Lemmer et al. [8] proposed four classes planetary habitats, those with: (1) surface water with conditions (e.g. moderate temperatures, etc.) to allow for multi-cellular life; (2) surface water allowing for organisms but not multi-cellularity; (3) subsurface water in contact with silicate rock bodies and magma; and (4) water bounded by water ice, or other fluids acting as substrate for life. The presence of these habitats depends on parameters of the host star. For example, M-class brown dwarf stars, making up about 75% of all stars in the Milky Way Galaxy, have a much narrower orbital band where planets could be potential hosts to liquid water—between 1/5 to 1/50 the width of that of systems hosted by G-class stars like the Earth's sun. [9] The presence of habitability also depends on planetary effects, and these may be so broad as to call into question the orbital-band-type calculations [10]. For example, if radionuclides are a source of energy, estimating their abundance in hot centers allows for a set of probabilistic calculations to modify orbital-distance calculations. See for example Lugaro et al. [11]. Habitability is also presumed as a function of the presence of redox gradients, chemical availability, planetary history, and others. The presence of chiral excess of organic compounds may be one of these additional parameters.

Astrobiologists are presently searching for evidence of living organisms in environments far removed from terrestrial biota. In particular, exploratory missions are looking for self-sustaining chemical systems that undergo Darwinian evolution, termed informally as the *NASA definition* of what constitutes a living organism [12]. Defining life is a difficult and controversial task. In Table 1, the NASA definition is analyzed using an assessment rubric, and it is seen that certain elements of what may define living organisms [13] are deemphasized, namely a boundary (e.g. a membrane) and heredity. It is a practical choice. Chemistry is measurable *in situ*, and modeling and experiments can be done from afar on Earth with consistent results to determine functional closure and evolutionary potential. Genetics are more difficult to establish, especially if a genetic system is unlike what is known. Organismal boundaries may also be difficult to establish with limited assessment equipment, i.e. remote microscopy and sample preparation may conflict with needs for sample preparation related to chemical assays—and the nature of what constitutes an acceptable boundary is not clear.

<Table 1>

To further elucidate the role of chiral excess in establishing habitability in astrobiological settings, a series of in-depth sections has been added to this introduction. These are arranged in the same order as the topics were introduced above: 1.1 Crystal Faces and Chirality, looking at mineral chirality and its influence on organic compounds; 1.2 Comets, Asteroids and Other Interstellar Matter, looking at processes affecting the creation of complex organic compounds and their chirality in source environments; 1.3 Magnetic Fields and Field Reversals, which explores a novel mechanism of creating chiral excess in near-stellar and near-planetary environments; and 1.4 Origin of Life: Overview, serving to provide greater basis for claims of an importance for chirality in biological systems. The methods section, results and discussion follow these.

**1.1 Crystal Faces and Chirality**

Structurally, there are left- and right-handed varieties of many minerals, and many of these can serve as templates for chiral enrichment via adhesion of organic molecules to the crystal face. Hazen [14] includes most common rock-forming minerals in this category: e.g. quartz, alkali feldspar, clinopyroxene and calcite. There are three patterns by which a crystal termination (i.e. a crystal face) may serve as a template for chiral enrichment of adhering molecules. It may do so if it has (1) a chiral planar tiling of atoms or groups of atoms on its surface; or (2) terminations of *achiral* sublattices that in their surface expression are chiral (i.e. edge effects); or (3) kinks which can serve as chiral centers. Lattice kinks due to impurities are ubiquitous in nature. Chiral enrichment can thus occur locally via templating on the faces of nearly all minerals. Of those minerals whose crystal lattices are achiral but whose terminations may support chiral enrichment of one *enantiomer* (i.e. chiral form) via template processes described above, calcite is perhaps the most important for its prevalence in aqueous settings in terrestrial environments [15]. Yet it is unclear whether the bulk result is chiral enrichment at a larger scale—without the addition of other factors.

Supposing that a template process is inherently important in chiral selection processes that occur *in situ* at life's origin, the following difficulty arises: In many environments, chiral forms of minerals are evenly distributed. Exceptions include: (1) local excesses, as shown statistically in some environments, e.g. due to seeding effects during mineral formation; (2) environments with structural anisotropy; (3) highly strained environments; and (4) environments with electrical or magnetic anisotropy.

Addressing (1) above: We know that as life on Earth and Earth itself coevolved, the diversity of minerals increased greatly [16]. To start at the beginning: None of the dozen minerals (i.e. diamond/lonsdaleite, graphite, moissanite, osbornite, nierite, rutile, corundum, spinel, hibonite, forsterite, perovskite, and silicate glass embedded with grains of metal and sulfides) found as inclusions in chondritic meteorites—and hypothesized to be remnant interstellar grains—belong to chiral symmetry groups. None of the primary minerals in chondrules of type 3.0 chondrites do either [16]. Yet interstellar grains of silicate glass, if they are large enough to be crystalline, i.e. quartz, will have a chiral crystal lattice. In fact, during the planetesimal stage of planetary formation where alteration and recrystallization of minerals occur, quartz is present [16]. Other chiral minerals (e.g. nepheline, clay minerals) formed during the earliest crustal evolution of the Earth. One sees that chiral minerals are present as soon as alteration and differentiation occur. Yet an even distribution of each enantiomer is assumed. Chiral excess would require another process, such as seeding, cf. Támara & Preston [17].

Addressing (2) above: Any depositional or directional process can produce structural anisotropy. Its presence does not specify a necessary planetary environment—myriad possibilities exist. Primary deposition or formation of minerals may be in layers due to settling or other processes assisted by gravity or other forces. Secondary alteration, likewise may take an *anisotropic* form—for example, equant mineral grains may dissolve and reprecipitate along structural interfaces over many successions.

Addressing (3) above: Strain can induce mineral-lattice changes via a variety of deformation processes, including brittle fracture, dissolution/precipitation, crystal plastic deformation, twinning and kinking, dynamic recrystallization, diffusion creep, granular flow, and others [18]. Notably, crystal plastic deformation can produce lattice preferred orientation (LPO) in an entire population of a mineral in response to state parameters, e.g. pressure, strain directions. Anisotropic fabric is found in mylonite zones, and in fact many of these oriented rock fabrics are features of deformation zones, such as one would find wherever tectonic activity occurs, e.g. in geothermal fields. Thus, if crystal templating is a necessary feature for a homochiral origin of living organisms, and if a necessary setting for this templating is in deformation zones (e.g. mylonites), then a history of planetary tectonism is a necessary feature of habitability if one assumes homochirality as essential to biology.

Addressing (4) above: Minerals that exhibit magnetic field interactions will orient along geomagnetic flux lines during their formation or deposition processes. At a planet that undergoes planetary magnetic reversals, such as Earth, one sees stripes of magnetic-field orientation of minerals in crustal environments symmetrical about spreading centers, i.e. mid-oceanic ridges [19]. Schematically, the process works as follows: First, emplacement of extrusive rock occurs over time during ongoing ridge building and extension, and magnetic minerals are aligned with the local magnetic field during cooling. Next, after a planetary magnetic reversal, new rock will exhibit minerals aligned in the new magnetic orientation. Finally, a series of reversals will produce magnetic stripes of opposite polarity symmetric about the ridge as extension of the ridge continues. Each stripe forms a record of the planetary magnetic field's orientation during the process of magma cooling. Orientation of magnetic minerals at the seafloor is not isotropic [19]. There is a systematic orientation to these minerals, and hence a plausible hypothetical source of chiral excess there from templating processes—wherein the magnetic anisotropy correlates with a local host environment favoring one chiral form of organic molecule over another [20]. A hypothesis follows: (a) If chiral templating on these minerals is a necessary feature for a homochiral origin of living organisms, and (b) if a necessary setting for this templating is at crustal zones adjacent to spreading centers (e.g. mid-ocean ridges), then (c) a history of planetary tectonism may be a necessary feature of habitability (d) if one assumes homochirality as essential to biology.

One may also consider examining minerals whose lattices exhibit anisotropic electrical (or magnetic) phenomena such as ferro- or paraelectricity. Anisotropic electrical fields can create conditions for chiral enrichment [21], and many common minerals and mineral families can produce such fields: alunite, apatite, beryl, cancrinite, epsomite, galena, ice, nepheline, prehnite, pyrochlore, quartz, rutile, serpentine, sodalite, sphalerite, topaz, tourmaline and zeolite—plus some meteorite minerals such as moissonite, rutile, perovskite and others [22]. Of these minerals that can produce electrical fields in the environment—and thus chiral enrichment—a few are of particular interest in a geologic setting that might be significant to the origin of life:

- Zeolites are porous minerals and occur in mid-ocean ridge environments, where the *chemical gardens* hypothesis of abiogenesis (i.e. the development of living organisms from inorganic matter) by Barge et al. [23] and others is favored. These can serve as substrates for cell-like structures.

- Galena is a sulfide mineral that is abundant in the mid-ocean ridge environment as well as others, such as geothermal fields, where abiogenesis may have occurred [24].

- Alunite and serpentine are secondary (alteration) minerals; they are significant if the origin of life occurs in weathered environments, i.e. where water was present.

- Apatite is of interest as a phosphate mineral, especially considering the importance of phosphorus for cellular metabolism and membranes.

- Ice is a prevalent mineral in many planetary, stellar-system, and interstellar bodies; it may act as a protonic pump in some occasions, as the most mobile charge carrier in ice is the proton rather than the electron.

- Minerals that produce electrical fields on meteorites, such as moissonite, rutile, perovskite, etc. are important as potential sources for chirally enriched organic compounds.

Yet what is described above is not a templating process, but rather one of mineral-induced field interactions with organic compounds during their formation or perhaps during a transportation or other process leading to chiral enrichment locally. Similar processes are described in the next section that treats phenomena occurring beyond the surface and atmosphere of a planet.

**1.2 Comets, Asteroids and Other Interstellar Matter**

One mode of abiogenesis involves the delivery of organic materials from interplanetary or interstellar sources. Once reaching a habitable environment, proto-life (materials important for prebiotic processes) are thought to evolve to living organisms. In support of this hypothesis, one may note that populations of amino acids exist on asteroids and comets [25,26]. It is reasonable to look for a mechanism (or mechanisms) related to its presence there, and also to mechanisms which could result in chiral enrichment.

The reactions in the dark clouds of interstellar dust at very low temperatures can be explained by quantum-mechanical tunnel effects (known as the Goldanski model or the *cold hypothesis*) which lead to chiral enrichment (also called *racemization*) of molecular materials [27]. A mechanism exists whereby racemization can occur through a stabilization process involving phonons (i.e. vibratory excess) at very low temperatures as is consistent with much of interstellar space [28].

Alternately, chirally enriched populations of organic molecules could likewise exist on bodies in this and other star systems in the galaxy, arising from polarized UV photons interacting with optically active materials that are chiral, including amino acids (called here the *warm hypothesis*). Meierhenrich et al. [29] demonstrated that right-polarized synchrotron radiation induced an enantiomer excess of D-leucine in solid state; the L- version was more readily broken down by the synchrotron radiation. Thus amino acids on (or in) meteorites and comets or in space generally could exhibit a chiral excess from high-energy, circularly polarized light extant in space.

During transmission, dust can polarize photons, as for example they pass through nebulae [30]. This photonic method of chiral selection is summarized as follows: (1) Photons pass through a dust cloud and become circularly polarized (CPL); and (2) CPL photons of sufficient energy interact with chiral (optically active) organic molecules, and the population of organic molecules becomes enriched in one enantiomer.

Thus, the distribution of dust may perhaps be a proxy for the likelihood of chiral enrichment of molecules that would interact with high-energy photons. Note that the distribution of organic molecules within interstellar dust is not uniform, thus the previous statement needs to be qualified. Reports that show the distribution of dust within the galaxy, such as that of Drimmel and Spergel [31] and others may be helpful in determining a likelihood of photonic polarization in different regions of the galaxy.

Finally note the following three points: (1) circular polarization due to cosmic microwave background (CMB) radiation is minute or beyond detection [32,33]; (2) pulsar (and other high energy cascade) magnetospheres are not a source of photons with circular polarization since the relevant contributions of positrons and electrons are in equal number and exactly cancel [34]; and (3) it is well known that white dwarfs are known CPL sources with strong magnetic fields, but the probability of exposure of organic molecules to this radiation is very low. Thus, galactic distributions of chiral excess due to CPL is likely limited to nebular interactions—creating distinct polarity in different galactic regions; and also to dust effects depending on local magnetic fields as planetary and other dusty systems develop.

Throughout the galaxy we may find regions of enantiomeric excess of organic molecules, but the chiral distribution of the biochemical machinery of living organisms may be approximately equal left- and right-handed overall. This presupposes the following:

- Racemization in complex organic molecules occurs primarily via CPL in dusty regions of interstellar space and in proto-planetary disks.
- The orientation of the magnetic fields in dusty regions of interstellar space and in proto-planetary disks is distributed equally in both of these environments overall. If not, there will be an excess due to magnetic field interactions with the dust. (See the following section for a more detailed discussion.)
- The likelihood of living organisms developing from the products of racemization in either of these two environments is equal. Othewise, local processes of interstellar space or proto-planetary disks will produce chiral enrichment that predominates—and then the distribution of racemization will follow the distribution of the environment that has the greater likelihood.

- The observed homochirality of amino acids (L-) and sugars (D-) in terrestrial life forms is not an essential feature of living organisms. If it is, then the observed distribution will not comprise a distribution of biochemical machinery, but only of enantiomeric excess.

Additional information about the process of racemization in proto-planetary disks is explored in the next section.

**1.3 Magnetic Fields and Field Reversals**

Magnetic fields can trap and influence the motion of dust particles via surface charging of those particles. This influence indicates a potentially important role for magnetic fields in the development of chirally abundant populations of optically active molecules, and hence of chiral organismal building blocks—if delivery of these materials is of interest. The following is a description of the process and a novel hypothesis about its mechanism in planetary settings. It is well-known that the magnetic fields of stars, and some planets, dwarf planets (e.g. perhaps Ceres) and satellites (e.g. Ganymede) reverse (oscillate) polarity over time. As these oscillations can influence the circular polarization of photons traveling through clouds of dust—affected also by magnetic fields in local environments—the oscillations affect chiral excess of organic compounds.

Stellar magnetic fields are caused by convective processes in plasma. A magnetic field reversal process occurs in stars as remanent magnetic flux in active regions (ARs) migrates poleward. It is a surface-flux process that is episodic, but subject to local inhomogeneities in the flux [35]. The above may also describe a hypothesis to explain the reversals of planetary magnetic fields. Instead of a geodynamo—with a solid inner core and liquid outer core—a different model would be adopted wherein the differential behaviors of seismic waves passing through the core's layers are due to differential properties in very dense plasmas. The Earth's inner core may be modeled as a high-density plasma. A list of key points in favor of generalizing this type of plasma process to planetary inner cores is given in Table 2. A *plasma-core model* is a simpler explanation for geomagnetic reversals than a geodynamo model, since the process is based on what is known of stellar magnetic reversals.

<Table 2>

The above description may be useful to estimate the occurrence of magnetic reversals in stars, some planets, dwarf planets and satellites in a galactic setting by modeling plasma processes. Periods of reversal could indicate a change in the predominant chirality of complex organic molecules within that environment. The energy in the field and proximity to the source need to be taken into account. The proposed process works as follows:

- Magnetic fields create surface charging of dust particles in the region.
- Photons interacting with these dust particles become circularly polarized (CPL).
- CPL photons of sufficient energy interact with chiral (optically active) organic molecules, and the population of organic molecules becomes enriched in one enantiomer.

Magnetic fields in stellar regions are extensive. Chiral excess is complicated by its potential relationship to magnetic reversals (and thus to stellar, and some planetary, dwarf planetary and satellite plasma processes). Calculating estimates of chiral enrichment in regions wherein magnetic oscillations would have occurred ought to provide a basis for testing the influence of this process. However, in planetary systems consisting of several bodies with different magnetic properties, this effect will not be always calculable because of the complex interaction between different magnetic fields.

**1.4 Origin of Life: Overview**

The chemical reactions important to life processes in organisms involve gradients. Metabolic processes make use of differences in temperature, pH, solute concentrations, electrical charge—separated by space, or by a boundary, in order to express localized results that favor desired outcomes. Environments wherein steep chemical gradients are found in a context of gradients in pH and temperature are candidate examples for an abiogenic origin of life.

The question of life's abiogenic origin is controversial. One hypothesis places the origin at thermal vents in *chemical gardens* [23,36,37]. Not only are there gradients in pH, temperature, and solute concentrations suitable to test for autocatalytic sets of chemistries that may evolve to include nucleic acids, but the rudiments of cell walls are also found, as, for example, an iron sulfide membrane [37] confirmed in the experiments of Barge et al. [23].

A competing hypothesis places life's abiogenic origin at the site of *geothermal fields*, where aliphatic compounds (cf. similar to soap scum) that come from meteoric sources interacts with and forms a membrane around organic chemical compounds [24]. Like at hydrothermal vents, geothermal fields also exhibit gradients in pH, temperature, and solute concentrations suitable for advantageous development of autocatalytic sets leading perhaps to selection and organismal origin.

A third hypothesis is developed by the author presently, building on similar work related to the foundations of metabolism in cold environments, i.e. the cold origins of life [38,39]. Cyclical changes in temperature as may be caused by orbital motion of icy bodies can provide additional chemical and physical gradients. Our novel contribution: The abiogenic origin of life may take advantage of low-temperature expression of electricity as a source of energy and change of pH from water ice—which can act as a source of protons [40]. Abiogenic origin from *ice substrates* requires the presence of additional complex molecules to form both membranes and the mechanics of heredity to take advantage of such gradients. Hence life's origin via ice might be restricted to locations with suitable mixtures. Note that the s*nowball Earth hypothesis* of terrestrial evolution [41] may lend credence to an ice-substrate hypothesis.

Notwithstanding which functional feature of organisms is more deeply or historically fundamental—whether membrane, metabolism, or heredity [13]—the presence of metabolic processes is inherent in organisms. The ability to transform chemical species from one molecule to another along a pathway of catalytic or enzymatic processes that bypass the slow kinetics of the general environment is very important to life. The following discussion highlights the importance of chirality in catalytic and enzymatic processes.

Enzymes and catalysts speed the kinetics of individual chemical reactions. The difference between an enzyme and a catalyst is nominal, i.e. enzymes are defined as organic, and typically these incorporate metal ion catalysts into their chemical structure [42]. The various forms of chlorophyll, for example, includes a magnesium ion at their center for the process of photosynthesis [43]. Broadly, enzymes and catalysts make use of molecular surface features and shape for their functionality.

In biochemical systems, many reactions are *homochiral*, i.e. they are able to utilize only one of two available enantiomers [44,45]. This is especially obvious for enzymes which themselves are proteins whose function relies on shape. Enzymes help with reaction kinetics and are so important to living organisms that the presence of a catalyst hosted within a molecular framework whose generation is encoded by another process (the author asserts) implies the presence of organisms. It is where one would look for evidence of foreign/novel types of organisms if there were no preconceived notions of how to proceed with a search. An outline of how enzymatic processes are central to the importance of chiral excess and homochirality in living organisms via abiogenesis is given below. Note that the outline progresses backwards from biological to physical processes.

- Metabolism relies on rapid chemical kinetics. For example: respiration needs to be rapid enough to provide energy to an organism.
- Rapid chemical kinetics rely on catalysts and enzymes. For example: glycolysis and pyruvate decarboxylation—precursory to the citric acid cycle during cellular respiration—rely on ten and one enzymes, respectively.
- Catalysts and enzymes rely on their physical shape for proper functioning. For example: hexokinase is the first enzyme of ten involved in glycolysis, and provides a platform for the phosphorylation due to its shape.
- Physical shape and chirality rely on physical phenomena following physical laws. For example: the shape of hexokinase is due to the bond energies of its constituent atoms.

In each of these instances, the reliance is essential for the proper functioning of the step above. The reliance on physical laws in the last step begs the question of whether the enantiomers used by living organisms in terrestrial environments are necessary to life. For enzymes, chirality affects activity. RNA acts as an enzyme on its own synthesis [46], thus some nucleic acids can be enzymatic as well as genetic. It is an open question.

2. Methods

A summary of findings from the ideas collected and developed in this article above can be found in Figure 1. These findings have been used to develop a worksheet with fillable fields that can serve to estimate the galactic distribution of sources of racemic complex organic compounds that could form living organisms in three different hypothesized origin of life sites. The file itself is given as a supplementary file, and also is discussed in the results and discussion sections below.

<Figure 1>

The algorithm used is very simple. An estimate of the relative importance and effectiveness of each feature identified is input by the user. Input values are from zero to 100, and then the results are normalized—so that the person entering the data need not calculate the totals (i.e. statistical population totals) in the dataset beforehand. The values can be a rough estimate, or can be based on external calculations of each process. The aim was to present a flexible tool to estimate relevant distributions of racemic organic compounds in different environments.

## 3. Results

The supplementary file contains three worksheets, and these are visible as the three tabs at the bottom of the file. One (named *Worksheet*) is the template without data entered. Another (named *with Arbitrary Values*) is a scenario with arbitrary data entered (i.e. a sequence of 10, 20, 30, ... 100, 0, 10 ... ) to check the functioning of the algorithms. This tab is also a reasonable location for a user to experiment with data values. Third, there is a tab (named *with Estimated Values*) that has values subjectively estimated for a prebiotic Earth environment.

The file itself is a .xlsx document created with the free software application LibreOffice and can be opened with any software that is compatible with Microsoft Excel files. Clicking on a field to highlight it will allow the user to change input values.  Clicking on the fields that show calculated values, e.g. "Organic fraction from hot cores" will bring the formula and cells used in the calculation into the editing bar above. The reader can change these relations as needed.

Estimates for the early terrestrial environment are notable and are explicitly assumptions: (1) contributions from hot cores are at that time already absent; (2) current flux of organics from the interstellar medium has been set at a low value; (3) racemization due to quantum effects in the diffuse interstellar medium, and due to CPL processes caused by the influence of astronomical objects on their nearby environments are minimal or nil; (4) racemization due to magnetic fields in the near-earth region are minimal but not zero; (5) layered sediment deposition (e.g. of clay minerals) is seen as the most important templating racemization process, and, in the worksheet, accounts for slightly more than half of mineral-template racemization due to the favorable electrical environment of clay mineral surfaces, with recrystallization (notably in calcite) accounting for the majority of the remainder—based on the abundance of slicate clays and calcite; and (6) clay minerals and zeolites are seen as the most important sites for racemization due to mineral electricity. Based on the above assumptions, the racemization effectiveness of the early terrestrial environment is about 25% in total in this worksheet, with approximately 10% effectiveness for templating processes, and 30% for mineral electrical and magnetic fields.  These estimates were based on user inputs of the relative importance of each material for the process overall and the abundance of each material.

The above is a naive model of the prebiotic terrestrial environment. A set of worksheets can be used for different regions of the galaxy, and each result used separately, or in conjunction with each other. The results will vary depending on the user. For example, a user might wish to model a single stellar system with this worksheet by copying the sheet to new tabs and input each with differential assumptions to compare how these interact.  The final result might be specified as a range, while each tab keeps track of the values for each assumption.  Larger models to encompass larger areas of the galaxy.

At this point, much of the needed data are not present, and only subjective estimates are possible—for indeed who can rule out chemical gardens in favor of geothermal fields, for example, as a valid locus of abiogenesis? There are no extant data (e.g. laboratory data or robust theoretical calculations) that are currently able to refine these estimates. Instead, users can generate estimates of prebiotic planetary or near-planetary environments to develop their own ranking between the different factors or see the ratio between their consequences. Thus a series of models can be compared, for example, and their interactions can take a form similar to what is done in stable isotope analysis—with different processes contributing to an observed outcome to account for intermediate values.

## 4. Discussion

The worksheet explicitly allows for estimation. It can be modified by readers as they see fit, and can help to conceptualize data to further understanding of abiogenesis and habitability. At a minimum, it can serve as a reminder of how little is known of the processes involved in abiogenesis. *Knowing what we don't know* is an important step along the way towards finding answers to seemingly insurmountable questions. The role of chirality in the origin of life is one of these—but no less challenging is quantifying the ability of various processes to perform racemization of organic compounds in far-reaching environments, for example, in the cold interstellar medium, in hot-core regions of nebulae, on mineral surfaces due to templating, or in the presence of electrical fields generated by minerals.

The focus of research on these many processes will likely narrow as more is learned about how each of the minimal features of an ideal first organism—metabolism, heredity, and a membrane—arise together. In the Origin of Life: Overview section, this work has focused on metabolism and the importance of enzymes and catalysts, and like the NASA definition described in the introduction, leans heavily towards a view where metabolism is important, i.e. a *metabolism-first* scenario for the origin of living organisms. There are also fruitful adventures to be had looking at *heredity-first* or *membrane-first* scenarios, or scenarios that have some features of these three, e.g. the *infra-biological* units of Szathmery [47], or the *dual-closure* concept of Jagers op Akkerhuis [48].

An astute reader may note that thermoelectric minerals have not been included in the electrical minerals listed, although there is a field in the worksheet where one can input additional data to include these. Also missing are detailed fields for magnetic minerals, e.g. piezomagnetics, and again these may be included in the field relating to magnetic minerals in general. Many of the experts in these areas are materials scientists working far afield from astrobiology or biochemistry—and a challenge here may be to create additional opportunities to import information and interest across field-specific boundaries.

Origin-of-life studies, like space science, holds out the promise of practical applications of work that originated as an attempt to elucidate larger theories. It is hoped that this worksheet, which leans heavily on conceptual facility, will inspire further research as users have a chance to see what information is needed to make sense of the bigger picture.

## 5. Summary

Chirality as a feature of terrestrial life suggests neither a special place for local development of homochirality nor for extra-terrestrial enrichment and delivery. Each seems to have some strengths. Chiral enrichment on asteroids or comets could have been promoted via photon polarization or preferential growth on ice mineral faces or both. Chiral enrichment *in situ* in chemical gardens or geothermal fields could have occurred via a variety of processes, including zeolite or sulfide surface templating. The possibilities were outlined in the article to clarify these scenarios and to allow for the construction of the electronic worksheet that accompanies this article in the supplementary files. Estimates of the prebiotic terrestrial environment are given in one of the worksheet scenarios as assumptions and can be used by the reader to develop their own ranking between the different factors or see the ratio between their consequences. Some important aspects in the field, e.g. symmetry breaking, chemical amplification of enantiomeric excess, were not discussed.

## Acknowledgements

This article was originally developed while the author was researching topics in astrobiology during affiliation with Prescott College and in correspondence with Dr. Richard Gordon and Dr. Alexei Sharov who were then editing a volume exploring the habitability of the universe before Earth. The research did not receive any specific grant from funding agencies in the public, commercial, or not-for-profit sectors. Love always to my parents.